\documentclass[letterpaper,twocolumn,prl,showpacs,superscriptaddress]{revtex4}
\usepackage{color}

\usepackage[dvipdfm]{graphicx}
\usepackage{amsmath}
\usepackage{times,yfonts}

\renewcommand{\Vec}[1]{\text{\boldmath $#1$}}
\newcommand{\gt}[1]{\mbox{\textfrak{#1}}}

\newcommand{\fig}[1]{Fig.\ \ref{#1}}
\newcommand{\gyroavg}[2]{\langle #1 \rangle_{\Vec{#2}}}
\newcommand{\kpar}{k_{\parallel}}
\newcommand{\kperp}{k_{\perp}}
\newcommand{\kperpc}{k_{\perp {\rm c}}}

\newcommand{\vpar}{v_{\parallel}}
\newcommand{\vperp}{v_{\perp}}
\newcommand{\vth}{v_{\rm th}}

\newcommand{\figsize}{7cm}

\newcommand{\Culham}{Euratom/UKAEA Fusion Association, Culham Science Centre, Abingdon OX14 3DB, United Kingdom}
\newcommand{\Oxford}{Rudolf Peierls Centre for Theoretical Physics, University of Oxford, Oxford OX1 3NP, United Kingdom}
\newcommand{\UMD}{Department of Physics, IREAP and CSCAMM, University of Maryland, College Park, Maryland 20742, USA}

\begin{document}

\title{
  Nonlinear phase mixing and phase-space cascade of entropy
  in gyrokinetic plasma turbulence
}

\author{T.~Tatsuno}
\author{W.~Dorland}
\affiliation{\UMD}

\author{A.~A.~Schekochihin}
\affiliation{\Oxford}

\author{G.~G.~Plunk}
\affiliation{\UMD}

\author{M.~Barnes}
\affiliation{\UMD}
\affiliation{\Oxford}
\affiliation{\Culham}

\author{S.~C.~Cowley}
\affiliation{\Culham}

\author{G.~G.~Howes}
\affiliation{Department of Physics and Astronomy, University of Iowa, Iowa City, Iowa 52242, USA}

\begin{abstract}
  Electrostatic turbulence in weakly collisional, magnetized plasma can be interpreted as a cascade of entropy in phase space, which is proposed as a universal mechanism for dissipation of energy in magnetized plasma turbulence.
  When the nonlinear decorrelation time at the scale of the thermal Larmor radius is shorter than the collision time, a broad spectrum of fluctuations at sub-Larmor scales is numerically found in velocity and position space, with theoretically predicted scalings.
  The results are important because they identify what is probably a universal
  Kolmogorov-like regime for kinetic turbulence;
  and because any physical process that produces fluctuations of the gyrophase-independent part of the distribution function may, via the entropy cascade,
  result in turbulent heating at a rate that increases with the fluctuation amplitude, but is independent of the collision frequency.
\end{abstract}

\pacs{52.30.Gz, 52.35.Ra, 52.65.Tt}

\maketitle

\paragraph{Introduction.}
Turbulence is inherently nonlinear and dynamically complicated.
In the general case, a broad spectrum of
fluctuations is excited, in both wave number and frequency.
For turbulent, magnetized plasma, the equations of magnetohydrodynamics
provide a pedagogically rich description of the dynamics.  However, for those turbulent
eddies whose parallel wavelengths (relative to the magnetic field) are
comparable to or smaller than the collisional mean free path and whose
perpendicular wavelengths are comparable to or smaller than the Larmor
radius of one of the constituent species of the plasma,
magnetohydrodynamic theory breaks down.
In such cases, the gyrokinetic (GK) theory
\cite{FriemanChen-PoF82,Howes-ApJ06}
represents a rigorous limit of plasma kinetics 
for anisotropic ($k_\parallel\ll k_\perp$),
low-frequency ($\omega\ll\Omega$, the ion cyclotron frequency) fluctuations. 
In this Letter, we present a GK description of turbulence in
a simplified situation, chosen to isolate a novel phenomenon which is
a generic component of all GK turbulence: the simultaneous
cascade of entropy to smaller scales in both real space and velocity
space. This phase-space cascade is the mechanism by which turbulent
energy associated with fluctuating fields is
brought to small scales in velocity space, where even very infrequent
collisions are sufficient to provide irreversibility and thus heating.
Below, we present the theory and first-principles simulations of the
phase-space cascade in a homogeneous, electrostatic, magnetized
plasma.

It is well known that Landau and Barnes damping of electromagnetic
plasma fluctuations lead to the generation of small-scale structures
in $f(v_\parallel)$, where $f$ is the one-particle distribution
function, and $v_\parallel$ is the velocity coordinate along the
background magnetic field \cite{Krommes,WatanabeSugama-PoP04}.
This is associated with the free-streaming of particles along
the field.
As $t$ increases,
a single Fourier harmonic of the distribution function
$f_{k_\parallel} \sim e^{i k_\parallel v_\parallel t}$ gets progressively more oscillatory in $\vpar$-space.
Eventually, even infrequent collisions are sufficient to smooth
these oscillatory features, since the collision operator is roughly
a diffusion operator in velocity space.
As long as collisions are sufficiently infrequent,
the damping rate depends not on the collision rate, but
 on the nature of the wave and its phase
velocity relative to the thermal speeds of the plasma species.
Physically, Landau damping is the smearing of spatial
perturbations that occurs when there is a spread in the distribution
of parallel velocities. We recall for future reference that this generation of
velocity-space structure is independent of the fluctuation
amplitudes. 

\begin{figure}
  \centerline{\includegraphics[width=5cm]{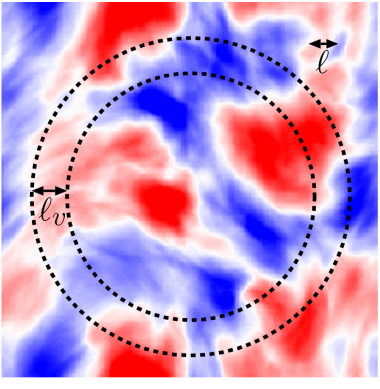}}
  \caption{
    Schematic view of the nonlinear phase mixing superimposed
    on the potential from the Run (iii) (see Table \ref{run table})
    at $t/\tau_{\rm init}=10$ and the largest wavelength mode taken out.
    When the fluctuation scale $\ell \lesssim \rho$, the gyroaverage of the 
    electric field induces
    a decorrelation of the distribution function at the 
    velocity-space scale corresponding to the difference in Larmor radii
    $\ell_v=\delta v/\Omega\sim\ell$ [see \eqref{correlation}].
  }
  \label{fig:mixing}
\end{figure}

Besides this {\em linear} parallel phase mixing,
there exists a {\em nonlinear} phase
mixing process \cite{DorlandHammett-PoFB93} 
that, in a strongly turbulent plasma and at spatial scales smaller than 
the Larmor radius, drives the formation of structure in $f(v_\perp)$
much more rapidly than parallel phase mixing drives $f(v_\parallel)$.
Physically, this nonlinear phase
mixing is the smearing of spatial perturbations 
due to the {\em spread} in the distribution of gyroaveraged $\Vec{E} \times \Vec{B}$ 
velocities (see \fig{fig:mixing}).  
Unlike for the parallel phase mixing, the rate of generation of
$v$-space structure by this process is proportional to the fluctuation
amplitude.
In this Letter, we present a study of this nonlinear process, which we 
interpret as a turbulent cascade of entropy in phase space \cite{Alex-tome}.
As such, it represents a conceptually novel nonlinear phenomenon, where 
generation of small scales in the position and velocity space occurs 
in an intertwined way.
This process, which is likely to be a fundamental 
and ubiquitous feature of magnetized plasma turbulence, has never been 
numerically diagnosed and analyzed before,
although Krommes \cite{Krommes} did point out the general possibility of the coupling between position and velocity space.

\paragraph{Gyrokinetics in 2D.}
Let the distribution function be $f=F_0 + \delta f$, where 
$F_0$ is a Maxwellian with density $n_0$ and temperature $T_0$, 
and $\delta f = h - q\varphi F_0/T_0$, where $q$ is the particle
charge and $\varphi$ is the electrostatic potential.
To keep the focus on the nonlinear process, we consider electrostatic 
GK turbulence in slab geometry with $k_\parallel = 0$.
Then the non-Boltzmann part $h$ of the perturbed ion distribution function satisfies \cite{FriemanChen-PoF82} 
\begin{equation}
  \label{eq:gk}
  \frac{\partial h}{\partial t} +  \frac{c \, {\Vec{\hat z}} \times \nabla
  \gyroavg{\varphi}{R}}{B_0} \cdot \nabla h = \gyroavg{C[h]}{R} 
  + \frac{q F_0}{T_0} \frac{\partial \gyroavg{\varphi}{R}}{\partial t},
\end{equation}
where $B_0$ is the background magnetic field aligned with the $z$-axis
and $\gyroavg{\cdot}{R}$ is the gyroaverage holding the
guiding center position $\Vec{R}$ constant. The collision
operator $C[h]$ used in our simulations contains pitch-angle scattering and energy
diffusion with proper conservation properties \cite{Abel}.
The quasineutrality condition yields
\begin{equation}
  \label{eq:qn}
  Q \varphi = q \int \gyroavg{h}{r} \, d\Vec{v}
  = q\sum_{\Vec{k}} e^{i\Vec{k}\cdot{\Vec{r}}} 
  \int J_0\left( \frac{k_\perp v_\perp}{\Omega}\right) h_{\Vec{k}}\, d\Vec{v},
\end{equation}
where $\gyroavg{\cdot}{r}$ denotes the gyroaverage at fixed
particle position $\Vec{r}$, 
$J_0$ is the Bessel function, $Q = \sum_s q_s^2 n_{0s} / T_{0s}$ for 
Boltzmann-response (3D) electrons or $Q = q_i^2 n_{0i} / T_{0i}$ for 
no-response (2D) electrons, and $s$ and $i$ are the species indices.
Our results are not affected by the choice of the electron response. 
For concreteness, we henceforth use no-response electrons since electrons cannot
contribute to the potential if $\kpar = 0$ exactly. 
In the absence of collisions, the system has two positive definite 
conserved integrals
\cite{Alex-tome,Plunk-JFM}:
\begin{align}
  \label{eq:wes}
  \gt{W} &= \iint \frac{T_0 \delta f^2}{2 F_0} \, d\Vec{r} \, d\Vec{v}
  = \int \left( \int \frac{T_0 \gyroavg{h^2}{r}}{2 F_0} \, d\Vec{v}
  - \frac{Q}{2} \varphi^2 \right) d\Vec{r}, \\
  \label{eq:w4d}
  \gt{E} &= \frac{Q}{2} \sum_{\Vec{k}} (1 - \Gamma_0) |\varphi_{\Vec{k}}|^2,
\end{align}
where $\Gamma_0 = I_0(\kperp^2\rho^2/2) e^{-\kperp^2\rho^2/2}$, $I_0$ is the modified Bessel function and $\rho$ is the ion thermal Larmor radius.
The invariant $\gt{W}$ is proportional to minus the perturbed part of the entropy of the system, $-\int f \ln f \, d\Vec{r} \, d\Vec{v}$ \cite{Krommes,WatanabeSugama-PoP04}. 
Here we will refer to $\gt{W}$ as ``entropy'' to emphasize this connection.
The second invariant $\gt{E}$ is conserved in the 2D electrostatic case only.

\begin{figure}
  \includegraphics[height=\figsize,angle=270]
    {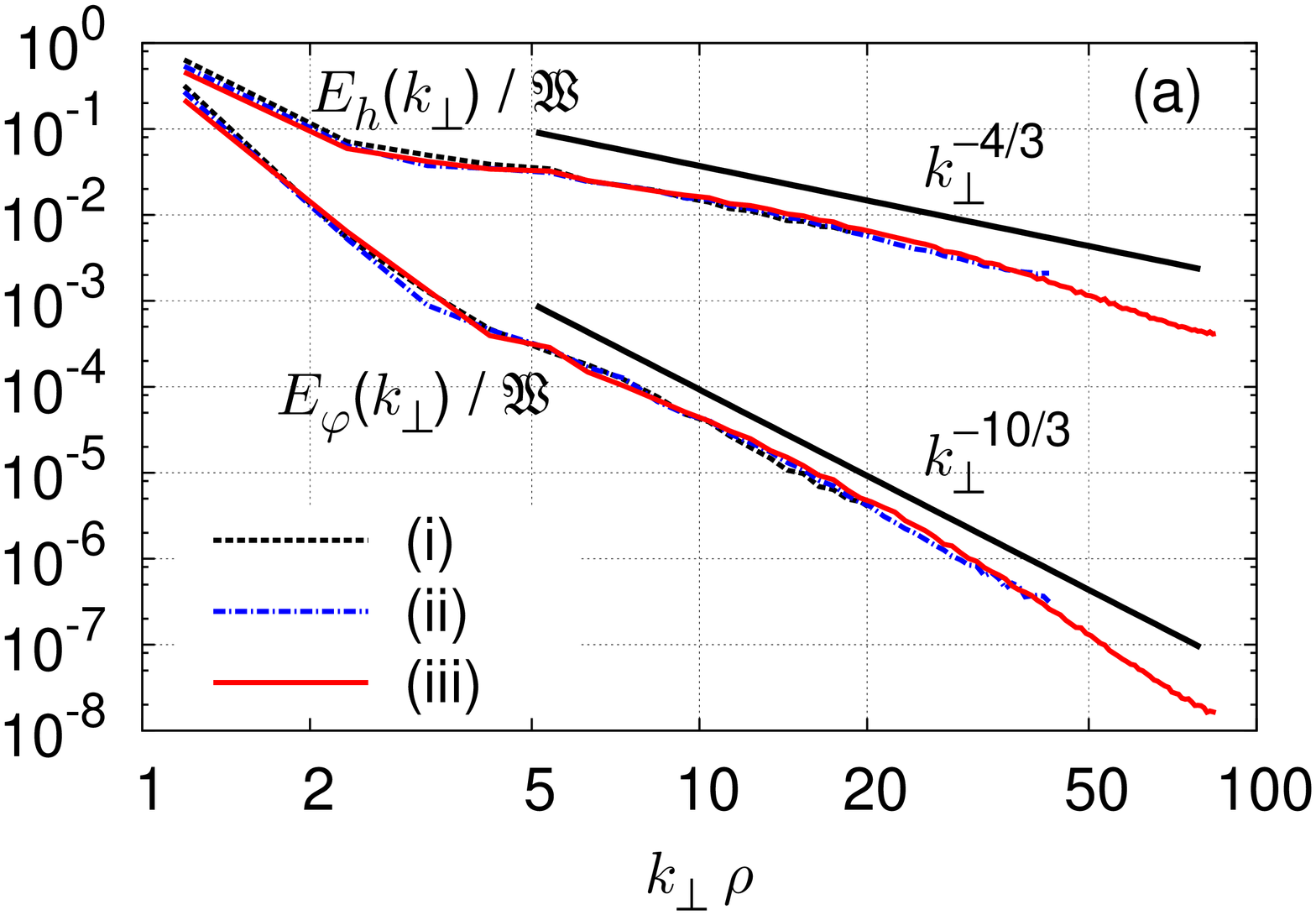}
  \includegraphics[height=\figsize,angle=270]
    {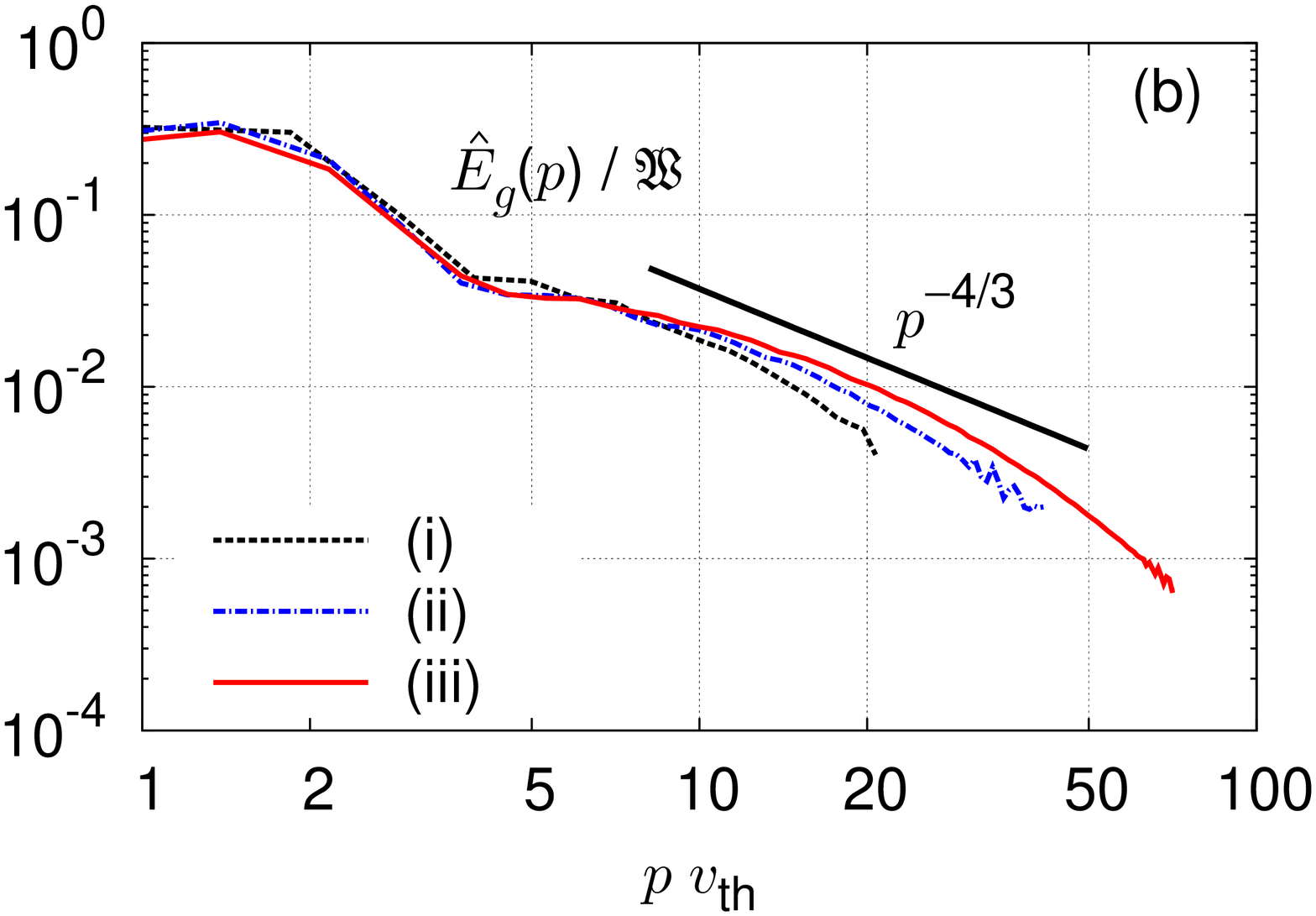}
  \caption{
    Time-averaged normalized (a) wave-number (Fourier) spectra $E_h(\kperp)/\gt{W}$ and $E_{\varphi}(\kperp)/\gt{W}$ [cf.\ \eqref{eq:spectra}], and (b) velocity-space (Hankel) spectrum $\hat{E}_g(p)/\gt{W}$ for the runs indexed in Table~\ref{run table}.
    Theoretically predicted slopes are given for comparison. 
  }
  \label{fig:spectra}
\end{figure}

\paragraph{Scalings.}
A scaling theory of the entropy cascade 
in the sub-Larmor scale range can be developed 
in a way reminiscent of the Kolmogorov-style turbulence theories \cite{Alex-tome}.
Assume that at (perpendicular) scales $\ell\ll\rho$, the transfer of entropy 
is local in scale. On dimensional grounds, the entropy flux is
\begin{equation}
  \frac{\vth^2}{\tau_{\ell}} \left( \frac{h \vth^3}{n_0} \right)^2
    = {\rm const}
    \label{entropy flux}
\end{equation}
until it reaches the collisional dissipation scale, where $\vth$ is the thermal speed and $\tau_{\ell}$ is the nonlinear decorrelation time at scale $\ell$.
The neglect of the $\varphi^2$ term in $\gt{W}$ [see \eqref{eq:wes}] is justified {\em post hoc} due to its smallness in the $\ell \ll \rho$ regime [see \eqref{eq:spectra} and \fig{fig:spectra}(a)].
There is a self-consistent electrostatic potential at the scale $\ell$:
from \eqref{eq:qn},
\begin{equation}
  \frac{q\varphi}{T_0} \sim \left(\frac{\ell}{\rho}\right)^{1/2}
    \frac{h \vth^3}{n_0} \left(\frac{\delta \vperp}{\vth}\right)^{1/2}
    \sim \frac{h \vth^3}{n_0} \frac{\ell}{\rho}.
  \label{QN dim}
\end{equation}
Here we have assumed that the nonlinear phase mixing produces velocity-space structures 
correlated with the spatial scale via 
(see \fig{fig:mixing} and Refs.~\cite{Alex-tome,Plunk-JFM})
\begin{equation}
  \frac{\delta \vperp}{\vth} \sim \frac{\ell}{\rho}.
  \label{correlation}
\end{equation}
This has allowed us to estimate the velocity integral in \eqref{eq:qn} 
as a random-walk-like accumulation of the integrand represented by the product 
of $h_{\Vec{k}}$, which is random function of $v_\perp$ whose ``step size'' is given 
by \eqref{correlation} with $\ell\sim k_\perp^{-1}$, and of the Bessel function,
which introduces a reduction factor of $(\ell/\rho)^{1/2}$.

The decorrelation time $\tau_{\ell}$ may be estimated by 
balancing the $\partial_t$ term with the nonlinear term in \eqref{eq:gk}, leading to
\begin{equation}
  \tau_{\ell} \sim \frac{\ell^2}{c\gyroavg{\varphi}{R}/B_0}
    \sim\left(\frac{\rho}{\ell}\right)^{1/2}
    \frac{\ell^2}{c\varphi / B_0}.
  \label{decorrelation time}
\end{equation}
Substituting \eqref{QN dim} into \eqref{decorrelation time} and \eqref{decorrelation time} 
into \eqref{entropy flux} yields $h \sim \ell^{1/6}$ and $\varphi \sim \ell^{7/6}$.
Therefore, the spectra of $h$ and $\varphi$ are 
\begin{equation}
  E_h(\kperp) \sim \kperp^{-4/3}, \quad E_{\varphi}(\kperp) \sim \kperp^{-10/3},
  \label{eq:spectra}
\end{equation}
where $E_h(\kperp) = \sum_{|\Vec{k}_\perp| = \kperp} \int T_{0} |h_{\Vec{k}}|^2 / 2F_{0} \, d\Vec{v}$ and $E_{\varphi}(\kperp) = \sum_{|\Vec{k}_\perp| = \kperp} q^2 n_{0} |\varphi_{\Vec{k}}|^2 / 2T_{0}$.
Note that the total entropy \eqref{eq:wes} can be expressed as $\gt{W} = \int [E_h(\kperp) - E_{\varphi}(\kperp)] \, d\kperp$.

\paragraph{Dissipation Cutoff.}
From the balance between the nonlinear decorrelation time \eqref{decorrelation time} and the collision time $\nu^{-1}$, one obtains an estimate of the dissipation cutoff scales in both the velocity space and the real space [see \eqref{correlation}].
Using $C[h] \sim \nu \vth^2 h / \delta v_\perp^2$, we find the cutoffs
\begin{equation}
  \frac{\delta v_{\perp{\rm c}}}{\vth} \sim \frac{1}{\kperpc\rho} \sim D^{-3/5},
  \quad  D = \frac{1}{\nu \tau_{\rho}},
  \label{cutoff scale}
\end{equation}
where $\tau_{\rho}$ is the nonlinear decorrelation time measured at $\ell = \rho$. 
We have introduced a new dimensionless number $D$ to characterize the scale separation in gyrokinetic turbulence:
analogous to the Reynolds number in fluid turbulence, large $D$ corresponds to a broader scaling range over which the entropy cascade extends, and to dissipation at smaller scales.
Here, however, the smallest spatial scale observed is determined by the $v$-space scale for which diffusion in velocities becomes important, through the correlation between real and velocity space given by \eqref{correlation}.
The fact that $D$ increases with the amplitude of the fluctuations at the Larmor scale clearly distinguishes this process from linear Landau damping.
We note that for 3D gyrokinetic turbulence, the nonlinear phase mixing is a much faster process than the linear one if the fluctuation amplitude is sufficiently large.

\paragraph{Numerical Simulations.}
We now report the first-of-a-kind numerical investigation of the entropy cascade in phase space, carried out with the GK code AstroGK.
The code uses a Fourier pseudo-spectral scheme for the real-space dimensions perpendicular to the background magnetic field and a Legendre collocation scheme for the velocity-space integrations.
The velocity space is discretized in energy $\varepsilon=v^2$ and $\lambda = \vperp^2/\varepsilon$. 
In the absence of collisions, AstroGK conserves the invariants \eqref{eq:wes} and \eqref{eq:w4d} with a high precision.

The results reported below were obtained in three runs at 
decreasing collision frequency $\nu$ and correspondingly 
increasing spatial and velocity resolution. They are indexed in Table~\ref{run table}, 
where $N_x\times N_y$ is number of collocation 
points in the real space and $N_\varepsilon\times 2 N_\lambda$ is the number of grid points 
in velocity space --- the factor of 2 corresponds to the sign of 
$\vpar = \pm\sqrt{\varepsilon(1-\lambda)}$. 
Our highest-resolved run required 36 wallclock hours on 8192 processors. 

\begin{table}
  \caption{Index of the runs.}
  \begin{center} \begin{tabular}{ccccccc}
      \hline \hline
      Run  & $N_x\times N_y$ & $N_\varepsilon\times 2N_{\lambda}$ & 
             $\nu \tau_{\rm init}$ && $D$  & $\kperpc\rho$ \\ \hline
      (i)  & $64^2$ & $32^2$ &
             $5.6 \cdot 10^{-3}$ && 48 & 20 \\
      (ii) & $128^2$ & $64^2$ & 
             $1.9 \cdot 10^{-3}$ && 118 & 35 \\
      (iii)  & $256^2$ & $128^2$ & 
             $7.4 \cdot 10^{-4}$ && 440 & 77 \\
      \hline \hline
    \end{tabular} \end{center}
  \label{run table}
\end{table}

\begin{figure} \begin{center}
    \includegraphics[height=\figsize,angle=270]
      {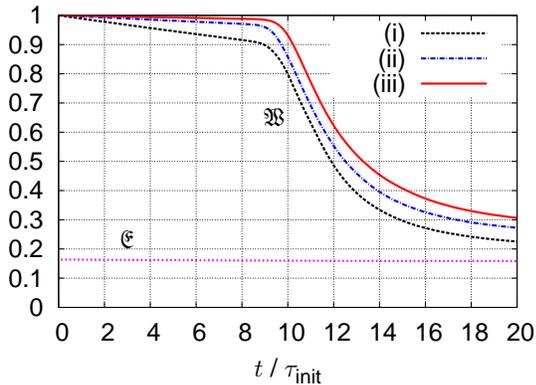}
  \end{center}
  \caption{
    Time evolution of $\gt{W}$ and $\gt{E}$ [Eqs.\ \eqref{eq:wes} and \eqref{eq:w4d}]
    normalized to initial $\gt{W}$.
    The runs (i)--(iii) are indexed in Table \ref{run table}.
    Evolution of $\gt{E}$ does not differ among runs significantly,
    and is given for run (iii).
  }
  \label{fig: time evolution}
\end{figure}

The code evolves $g = h - q F_0 \gyroavg{\varphi}{R} / T_0$ and $\varphi$ via 
Eqs.~\eqref{eq:gk} and \eqref{eq:qn}.
We take the box size $L_x = L_y = 2 \pi \rho$ and start from the initial condition $g_{\rm init} = g_0 [ \cos (2x/\rho) + \cos (2y/\rho) + \chi(x,y) ] F_0$, where $g_0$ is a constant and 
$\chi(x,y)$ is a small-amplitude white noise superimposed on all Fourier modes.
From \eqref{eq:qn}, we can calculate $\varphi_{\rm init}$.

\paragraph{Time Evolution.}
The initial $|k_x\rho|, |k_y\rho|=2$ configuration is unstable:
the amplitudes of $\varphi$ corresponding to $|k_x\rho|, |k_y\rho|=1$ grow 
and then saturate around $t/\tau_{\rm init} \simeq 9$, where $\tau_{\rm init} = 2\pi B_0 / (c \kperp^2 ||\gyroavg{\varphi_{\rm init}}{R}||)$ is the turnover time associated with the initial condition and $||\gyroavg{\varphi}{R}|| = [(1/n_0) \iint |\gyroavg{\varphi}{R}|^2 F_0 \, d\Vec{v} \, d\Vec{R}]^{1/2}$.
The nonlinear interactions between modes produce smaller scales down to a cutoff determined by $D$ [see \eqref{cutoff scale}].
The turbulent spectra fill up by $t/\tau_{\rm init} \simeq 10$, then decay with time. 

The time evolution of the collisionless conserved quantities $\gt{W}$ and $\gt{E}$ 
[see \eqref{eq:wes} and \eqref{eq:w4d}] is shown in \fig{fig: time evolution}.
During the initial growth of the instability of the $|k_x\rho|, |k_y\rho|=1$, 
$\gt{W}$ decays very slowly at a rate $\sim\nu$, consistent with a collisional 
decay rate associated with the large-scale phase-space variation of $g_{\rm init}$. 
Once turbulence develops, $\gt{W}$ decays more rapidly as the entropy cascade 
transfers it nonlinearly to smaller scales in phase space, until the fluctuations 
of the distribution function are thermalized (dissipated) at the collisional cutoff.

The decrease of $\gt{W}$ from its initial value corresponds to the amount of entropy (heat) production due to the irreversible collisional smearing of the distribution function. 
The turbulence that follows the initial instability enhances the heating, suggesting that small-scale velocity-space structure is generated (this is confirmed below). 
As expected, the rate of dissipation is not strongly affected by the collision frequency, i.e., there is a finite amount of dissipation even as the collision frequency tends to zero. The dissipation rate is determined instead by the nonlinear cascade rate. 

While $\gt{W}$ decays, $\gt{E}$ stays almost constant.
If we increase the size of the simulation box, the $|k_x\rho|, |k_y\rho|=1$ modes 
themselves become unstable to even longer-wavelength modes. We attribute both this instability and the failure of $\gt{E}$ to decay to the intrinsic tendency of $\gt{E}$ to have an inverse cascade \cite{Idomura,Plunk-JFM}, which we do not discuss here.

\paragraph{Spectra and Scalings.}
The wave-number spectra of the decaying developed turbulence are given in \fig{fig:spectra}(a).
They are angle integrated over wave-number shells $|\Vec{k}_\perp|=\kperp$, 
normalized by $\gt{W}(t)$ at each time and then averaged 
over time for $10 \le t/\tau_{\rm init} \le 15$. 
As resolution is increased, 
the spectra appear to converge to the theoretically predicted scalings \eqref{eq:spectra}, 
which supports the validity of our dimensional and physical considerations of the entropy cascade.

To characterize the entropy cascade in the velocity space, 
Plunk \textit{et al}.~\cite{Plunk-JFM} introduced velocity-space spectra 
$\hat{E}_g(p) = \sum_{\Vec{k}} p |\hat g_{\Vec{k}}(p)|^2$,
where $\hat g_{\Vec k}(p) = \int J_0(p \vperp) g_{\Vec{k}}(\Vec{v}) \, d\Vec{v}$ is a Hankel transform.
The theoretical expectation is that $\hat{E}_g(p) \sim p^{-4/3}$ because the real- and velocity-space scales should be related according to \eqref{correlation}, which, in terms of the dual variable $p$, becomes $\kperp\rho\sim p\vth$.
The time-averaged Hankel spectrum $\hat{E}_g(p)$ obtained in our simulations is shown in \fig{fig:spectra}(b).
This again shows approximate consistency with the theoretical prediction and confirms that small-scale structure is formed in the velocity space.

\paragraph{Dissipation Cutoff.}
In Table \ref{run table}, we show for each of our runs 
the dimensionless number $D=(\nu\tau_\rho)^{-1}$, 
where $\tau_{\rho} = 2\pi B_0/(c \kperp^2 ||\gyroavg{\varphi'}{R}||)$ measured at $t/\tau_{\rm init} = 10$ and $\varphi'$ is $\varphi$ with the $|k_x \rho|, |k_y \rho| = 1$ modes taken out (see also \fig{fig:mixing}). Also shown is 
the theoretical estimate \eqref{cutoff scale} for the wave-number cutoff $\kperpc\rho = \alpha D^{3/5}$, where $\alpha = 2$ is an empirical value that corresponds to our particular set up.
Comparing with the wave-number and velocity-space spectra in \fig{fig:spectra}, we see that \eqref{cutoff scale} describes the resolution requirements quite well.
With fewer velocity grid points, we find shallower wave-number spectra than the resolved ones, while with more, we resolve below the velocity cutoff without any change in the wave-number spectra.
Thus $D$ is a good indicator of necessary and sufficient resolution in full 4D phase space.

\paragraph{Conclusions.}
We have presented electrostatic, decaying turbulence simulations for weakly collisional, magnetized plasmas using the gyrokinetic model in 4D phase space (two real-space and two velocity-space dimensions). Landau damping was removed from the system by ignoring variation along the background magnetic field.
Nonlinear interactions introduce an amplitude-dependent perpendicular phase mixing of the gyrophase-independent part of the perturbed distribution function and create structure in $v_\perp$ which is finer for higher $\kperp$. 
We have found that the wave-number (Fourier) 
and velocity-space (Hankel)
spectra of the perturbed distribution function and the resulting electrostatic fluctuations at sub-Larmor scales agree well with theoretical predictions based on the interpretation of the nonlinear phase mixing as a cascade of entropy in phase space \cite{Alex-tome,Plunk-JFM}.
We have introduced a dimensionless number $D$ (analogous to Reynolds number) that characterizes the scale separation between the thermal Larmor scale and the collisional cutoff in phase space [see \eqref{cutoff scale}], and showed that this number correctly predicts the resolution requirements for our simulations.

We note that there are, in general, entropy cascades for each plasma species.
Equations for the gyrokinetic turbulence at and below the electron Larmor scale 
are mathematically similar to the model simulated here and identical arguments apply 
\cite{Alex-tome,Plunk-JFM}. Similar considerations are also possible for ion-scale electromagnetic turbulence \cite{Alex-tome} and for minority species. 

The small-scale phase-space structure that we have discovered is likely to be a universal feature of strong, magnetized plasma turbulence.
Understanding it theoretically and diagnosing it numerically is akin to the inertial-range studies for Kolmogorov turbulence, extended to the kinetic phase space.
One should expect rich and interesting physics to emerge and it is likely that predicting large-scale dynamics will require effective models for the small-scale cascade.
An immediate key physical implication of the existence of the entropy cascade is a turbulent heating rate independent of collisionality in weakly collisional plasmas.


Numerical computations were performed at NERSC, NCSA and TACC.
This work was supported by the U.S.\ DOE Center for Multiscale Plasma Dynamics, 
STFC, and the Leverhulme Trust Network for Magnetised Plasma Turbulence.

\end{document}